\documentclass[preprint,12pt,showpacs]{revtex4} 
\usepackage{amsmath}
\usepackage{latexsym}
\usepackage{amssymb}
\usepackage{graphics,epstopdf}
\usepackage{amsmath,amssymb,amsthm}
\usepackage{slashed}
\usepackage{graphicx}
\usepackage[colorlinks=true, citecolor=blue, urlcolor=blue ]{hyperref}
\theoremstyle{plain}

\begin{document}
\title{Dynamics of the free time-dependent effective mass}

\author{Pinaki Patra}
\thanks{Corresponding author}
\email{monk.ju@gmail.com}
\affiliation{Department of Physics, Brahmananda Keshab Chandra College, Kolkata, India-700108}

\author{Aditi Chowdhury}
\email{aditichowdhury01@gmail.com}
\affiliation{Department of Physics, Acharya Prafulla Chandra College, Kolkata, India-700131}

\author{Milan Jana}
\email{milanepsilon@gmail.com}
\affiliation{Department of Physics, Acharya Prafulla Chandra College, Kolkata, India-700131}

\date{\today}

\begin{abstract}
The consensus is that an object with a large mass will not manifest quantum behavior. Therefore, we expect that the quantumness of a  time-dependent effective mass (TDEM) will erase after a long time when the mass profile continuously grows with time. However,  the present article depicts that the Wigner quasi-probability distribution (WQD) will manifest an entanglement behavior forever for two spatially separated free TDEMs. \\
The time-dependent Schr\"{o}dinger equation for a free particle with TDEM has been solved with the help of the Lewis-Riesenfeld phase space invariant method. WQD for the system of two identical TDEMs with quadratically increasing mass profiles shows that the particles are never separated. In particular, their reminiscent is present at the origin of the phase-space forever. 
\end{abstract}
 
 \maketitle
\section{Introduction}
The idea of the existence of ideal free particles (FP) is rooted in the philosophy of Descartes, who had prescribed eliminating \textquotedblleft the influences from far away\textquotedblright   to remove the plague of medieval sciences \cite{Descartes}. Modern science adopted the concepts of the existence of free particles in a similar fashion \cite{Rovelli}. The notion of FP is widely accepted to be fruitful for the approximate behavior of a physical system. However, the quantum entanglement phenomenon seems to defy the strict meaning of FP \cite{Horodeci}. Therefore, it seems reasonable to claim that the problem of  FP deserves attention in its own right. In this article, we have revisited the aspects of a time-dependent effective mass, which is moving as an FP. 
The time-dependent effective mass (TDEM) appears to be relevant for a diverse domain of Physics, namely in the energy density functional approach to many-body problems \cite{Buranco1,Arias1}, in the study of electronic properties of condensed matter systems \cite{CostaFilho,Muharimousavi,Souza Dutra,Souza Dutra1,Schmidt,heterostructure1,heterostructure2,Mario}, the Schr\"{o}dinger equation in curved space under the shed of deformed algebras \cite{Quense}, nonlinear optical properties in quantum well \cite{optical properties,optical properties2}, and even the cosmological models to quantum information theory \cite{qinfo1,qinfo2,qinfo3,qinfo4,qinfo5,qinfo6} and many more \cite{santos,cavalcanti,cunha,bekke,vitoria,vitoria1,bekke1,delta1}. For instance,  the \textquotedblleft asymmetric shape of crackling noise pulses emitted by a diverse range of noisy systems\textquotedblright is modeled with the help of TDEM in \cite{crackling noise,crackling2}. For a TDEM, the system becomes non-conservative \cite{tdm1,tdm2}. In particular, if a quantum system interacts with time-varying environments such as temperature, pressure, stress, and energy, the effective masses will be modified by some TDEM \cite{tdm1,tdm2,tdm3,tdm4,tdm5,tdm6}.
A first principle calculation for the viable Hamiltonian of a position and time-dependent effective mass was proposed in \cite{pdem}. 
The non-perturbative solution of the time-dependent Schr\"{o}dinger equation (TDSE) for non-conservative systems (in particular for TDEM) is a tricky one (if not impossible). However, if a class of phase-space invariant operators (PSIO) corresponding to the time-dependent Hamiltonian ($\hat{H}(t)$) exists, then the system is exactly solvable. For example, the quadratic (both in position and momentum) Hamiltonians are solvable by some factorization method \cite{bilinear1,bilinear2,bilinear3}.\\
Lewis-Riesenfeld phase-space invariant method (LRIM) is of the classic formalisms for solving a TDSE with time-dependent parameters (e.g., mass $m(t)$) \cite{Lewis,Lewis1,Lewis2,Lewis3,Lewis4,Lewis5,Lewis6,Lewis7}. The idea behind the LRIM is to construct a PSIO corresponding to the Hamiltonian $\hat{H}(t)$. Up to a time-dependent phase factor, the
eigenstates of the PSIO ( $\hat{\mathcal{I}}$) will satisfy the TDSE corresponding to $\hat{H}(t)$ \cite{Penna,Bagrov,Torre,Guerrero1,Geloun,Maamache,Pinaki1,Ponte}. 
LRIM for various classes of TDSE had been studied extensively in the literature. Such as the harmonic oscillators, a particle moving in a time-dependent electromagnetic field, and a particle moving in non-commutative (NC) space with time-dependent NC parameters have been solved with the help of LRIM \cite{lrimapplication1,lrimapplication2,lrimapplication3,lrimapplication4,lrimapplication5}. 
M Maamache et al. constructed a class of Lewis-Riesenfeld invariant operators (LRIO) for a free particle with TDEM \cite{Maamache}. 
However, there is a gap in the literature on the general study of LRIO for an FP with TDEM. The present paper aims to fulfill this gap. The construction of the most general quadratic LRIO for an FP with TDEM is one of the motivations behind the present article.  
 Moreover, the restrictions on the
parameters of the LRIO are determined so that the diagonalization of $\hat{\mathcal{I}}$ with the help of a similarity transformation (symplectic group $sp(2,\mathbb{R})$) is possible. It turns out that, in diagonal representation, $\hat{\mathcal{I}}$  can be factorized in terms of annihilation ($\hat{a}$) and creation ($\hat{a}^\dagger$) operators. The eigenstates along with the corresponding eigenvalues of $\hat{\mathcal{I}}$ are then determined from the eigenstates of $\hat{a}$.  To obtain the complete solution of TDSE, we have computed the time-dependent phase factor (both the geometrical and dynamical phases).  With the help of the uncertainty relation in $\hat{x}$ and $\hat{p}$, we have shown that the ground state is a squeezed coherent state (CS).
Being the minimum uncertainty state, a CS mostly resembles the classical states \cite{wigner1,wigner2,wigner3,wigner4}. On the other hand, the Wigner quasiprobability distributions (WQD) provide a phase-space representation of QM. it is customary to study the Wigner quasiprobability distributions (WQD) for CS. WQD is generally felt to offer several advantages for use in modeling the behavior of physical processes. For instance, being a phase-space formulation, WQD involves both real space and the momentum space variables, distinctly different from SE \cite{wigner5,wigner6}.  It is believed that one can conceptually
identify where quantum corrections enter a problem by comparing it with the classical version. However, philosophical debates in this regard are inevitable, which we shall avoid in the present article. Rather, we shall indulge ourselves in the computation of WQD for our
problem, keeping in mind that, being entirely real, WQD simplifies both the calculation and the interpretation of results. For this reason, it was a natural choice for the simulation of quantum transport in devices such as the resonant tunneling diode \cite{wigner5,wigner6}.\\
In the present paper, we have considered a system of two noninteracting free particles with
TDEM, moving in the opposite direction to each another. For the constant mass case, it is well known that this type of bipartite state will produce an entanglement term at the origin of the phase-space \cite{wigner5,wigner6,wigner7}. Since amplitude depends on the mass, it is expected that the entanglement term will be diminished as time flows. However, we have shown that the amplitude of WQD remains intact at the origin of the phase-space.
The organization of the article is as follows. At first, a class of quadratic LRIO corresponding to a free particle with TDEM is constructed. Then the eigenstates and corresponding eigenvalues of the invariant operator are constructed with the help of the factorization method. The geometric and dynamic phase factors are determined in a closed form, which enables us to write down the exact solutions of TDSE corresponding to our system. It is shown that the states are indeed squeezed CS. Finally, we have constructed the WQD, which corresponds to the bipartite CS.

\section{Lewis-Risenfeld Invariant operator}
Lewis-Riesenfeld (LR) theorem \cite{Ponte} states that for a system described by a time-dependent (TD) Hamiltonian $\hat{H}(t)$, a particular solution of the associated TD Schr\"{o}dinger equation (SE), 
\begin{equation}\label{schrodinger}
 \hat{H}\psi =i\hbar \frac{\partial\psi}{\partial t},
\end{equation}
is given by the eigenstate $\vert n,t\rangle $ of a TD invariant $\mathcal{\hat{I}}$ defined by the equation
\begin{equation}\label{LRIdefn}
\frac{d \hat{\mathcal{I}}}{dt}= \frac{\partial  \hat{\mathcal{I}}}{\partial t} + \frac{1}{i \hbar}[ \hat{\mathcal{I}}(t),\hat{H}(t)] =0,
\end{equation}
apart from a TD phase factor $e^{i\theta(t)}$. Assuming the eigenvalue equation $\hat{\mathcal{I}}(t)\vert n,t \rangle = \lambda_n \vert n,t \rangle$ for a discrete spectrum, $n=0,1,2,...$, one can verify that $\dot{\lambda}=0$, and 
\begin{equation}\label{thetandot}
 \hbar\theta_n =\int_0^t\langle n,\tau \vert (i\hbar \frac{\partial}{\partial t}-\hat{H}(t))\vert n,\tau\rangle d\tau.
\end{equation}
Here dot denotes the derivative with respect to time. 
The general solution of the TDSE is given by the superposition state
\begin{equation}
 \vert \psi (t)\rangle = \sum_{n} c_n e^{i\theta_n} \vert n,t\rangle, \; \mbox{with}\; c_n= \langle n \vert \psi(0)\rangle.
\end{equation}
 We shall apply the LR theorem for the following Hamiltonian of a free particle.
\begin{equation}\label{hamiltonian}
 \hat{H}(t)=\frac{\hat{p}^2}{2m(t)}.
\end{equation}
At first, we observe that the set of operators
$
 \mathcal{A} =\left\{\hat{p}^2, \hat{x}^2, \left\{ \hat{x},\hat{p}\right\}\right\}
$
forms a closed quasi-algebra with respect to $\hat{H}$ of ~\eqref{hamiltonian}. 
In particular,
\begin{eqnarray}\label{quasialgebra}
[\hat{H},\hat{p}^2] = 0,\; [\hat{H}, \hat{x}^2] = -\frac{i\hbar}{m} \{\hat{x},\hat{p}\}, \;
[\hat{H}, \{ \hat{x},\hat{p}\} ] = -\frac{2i\hbar}{m}\hat{p}^2.
\end{eqnarray}
The algebraic relations ~\eqref{quasialgebra} suggest the following ansatz for the LR-invariant operator.
\begin{eqnarray}\label{ansatz}
 \hat{\mathcal{I}}(t) = \alpha(t) \hat{p}^2 + \beta(t)\hat{x}^2 + \gamma (t)\left\{\hat{x},\hat{p}\right\} +\delta(t)\hat{\mathbb{I}},
\end{eqnarray}
where $\hat{\mathbb{I}}$ is the identity operator.
Using ~\eqref{ansatz} and ~\eqref{hamiltonian} in~\eqref{LRIdefn}, 
we get the following set of first order coupled differential equations.
\begin{eqnarray}
\dot{\beta}&=&
\dot{\delta}=0. \label{betadotdeltadot}\\
\dot{\gamma}&=&-\frac{1}{m}\beta. \label{gammadot}\\
\dot{\alpha}&=&-\frac{2}{m}\gamma. \label{alphadot}
\end{eqnarray}
 According to equation ~\eqref{betadotdeltadot}, $\beta$ and $\delta$ are constants. Since, $\delta$ is the constant coefficient of the identity operator, without loss of generality, we can set it to zero. Moreover, using equations ~\eqref{gammadot} and ~\eqref{alphadot}, we can see that 
 \begin{equation}\label{kappasquare}
  \frac{d}{dt}(\gamma^2 - \beta \alpha)=0 \implies \gamma^2 - \beta \alpha = -\kappa_0^2,
 \end{equation}
 where $\kappa_0^2$ is a real constant.
Solving equations ~\eqref{gammadot} and ~\eqref{alphadot}, we get
\begin{eqnarray}
\gamma(t)&=& \gamma_0 -\beta \mu(t), \label{gammat}\\
\alpha(t)&=& \alpha_0 -2\gamma_0 \mu(t)+\beta \mu^2(t) , \label{alphat} \\
&& \mbox{with} \; \dot{\mu}=\frac{1}{m(t)}.
\end{eqnarray}
 $\alpha_0$ and  $\gamma_0$ are integration constants, subject to the constraint (using equation~\eqref{kappasquare}) 
 \begin{equation}
  \gamma_0^2 - \beta \alpha_0 = -\kappa_0^2.
 \end{equation}
For future convenience, let us define an auxiliary real-valued function $ \alpha(t)=\sigma^2(t)$.
Then, the system  ~\eqref{gammadot}-~\eqref{alphadot}  is equivalent to the auxiliary equation
\begin{equation}\label{gammasigma}
 \dot{\sigma}^2 = \dot{\mu}^2 (\beta + \kappa_0^2\sigma^{-2}).
\end{equation}
In terms of a real solution of $\sigma(t)$ ~\eqref{gammasigma}, the  invariant operator~\eqref{ansatz} takes the form
\begin{equation}\label{Iinsigma}
 \hat{\mathcal{I}}= \sigma^2 \hat{p}^2 + \beta \hat{x}^2 + \sqrt{\beta\sigma^2 -\kappa_0^2}\{\hat{x},\hat{p}\}.
\end{equation}

\section{Constraint on parameters for quadratic form}
If the integration constants in~\eqref{alphat} and ~\eqref{gammat} are considered to be zero ($\alpha_0=\gamma_0=0$), then ~\eqref{Iinsigma} can be written  in the quadratic form
\begin{equation}\label{quadraticI}
 \hat{\mathcal{I}}=\frac{1}{2}X^T \hat{\mathcal{H}}_0X,
\end{equation}
where
\begin{eqnarray}
 \hat{\mathcal{H}}_0=2\beta \left(\begin{array}{cc}
                                   1 & -\mu \\
                                   -\mu & \mu^2
                                  \end{array}
\right), \; X=(\hat{x},\hat{p})^T.
\end{eqnarray}
One can identify the intrinsic symplectic structure (symplectic group Sp$(2,\mathbb{R})$) 
\begin{eqnarray}\label{symplectics}
\left[\hat{\mathcal{I}}, X\right] = -i \hbar\hat{\Omega}X = \hbar \Sigma_y \hat{\mathcal{H}}_0 X,
\end{eqnarray}
with
\begin{eqnarray}\label{Omega}
 \hat{\Omega}=2\beta \left( \begin{array}{cc}
                                   -\mu & \mu^2 \\
                                   -1 & \mu
                                  \end{array}\right),\;
 \Sigma_y =\left( \begin{array}{cc}
    0 & -i \\
i & 0
\end{array}\right).
\end{eqnarray}
To have an equivalent coordinate system, in which $\hat{\mathcal{I}} $ is diagonalized, we have to construct a similarity transformation (symplectic group Sp($2,\mathbb{R}$)). The characteristic polynomial ($P_\lambda= Det(\hat{\Omega}-\lambda \hat{\mathbb{I}})$) of $\hat{\Omega}$,  has trivial roots $\lambda=0,0$. That means  $\hat{\mathcal{I}}$ can not be diagonalized by a canonical transformation keeping the Sp($2,\mathbb{R}$) structure intact for the parameter values $ \alpha_0= \gamma_0= 0$.
However, we can diagonalize the system for the parameter values
$\alpha_0\neq 0,\; \gamma_0\neq 0,\; \kappa_0\neq 0$, which is considered throughout this paper.
\section{Diagonalization of $\hat{\mathcal{I}}$ and evaluation of eigenstates}
By direct observation, it is not difficult to rewrite ~\eqref{Iinsigma} as 
\begin{eqnarray}\label{invariantcanonical}
 \hat{\mathcal{I}} = \left(\sigma\hat{p}-m\dot{\sigma}\hat{x}\right)^2+\kappa^2x^2 ,\;\;
 \mbox{with}\;\; \kappa^2(t)= \frac{\kappa_0^2}{\alpha(t)}.
\end{eqnarray}
~\eqref{invariantcanonical} suggests the following form for the annihilation operator $\hat{a}$ and the corresponding creation operator $\hat{a}^\dagger$.
\begin{eqnarray}
 \hat{a} = \frac{1}{\sqrt{\hbar\omega}}\left( \sigma \hat{p} -m\dot{\sigma}\hat{x}-i\kappa \hat{x}\right),\\
 \hat{a}^\dagger = \frac{1}{\sqrt{\hbar\omega}}\left( \sigma \hat{p} -m\dot{\sigma}\hat{x}+i\kappa \hat{x}\right).
\end{eqnarray}
The constraint 
\begin{equation}
 [\hat{a},\hat{a}^\dagger]=1 \; \implies \omega = 2\kappa_0.
\end{equation}
We can factorize  ~\eqref{invariantcanonical}  straightforwardly as
\begin{equation}\label{Iinaadagger}
 \hat{\mathcal{I}}= \left(\hat{a}^\dagger \hat{a} +\frac{1}{2}\right)\hbar\omega.
\end{equation}
$\hat{a}$ and $\hat{a}^\dagger$ act on the eigenstates ($\vert n\rangle$) of $\hat{\mathcal{I}}$ as follows.
\begin{eqnarray}
 \hat{a}\vert n\rangle &=& \sqrt{n}\vert n-1\rangle, \;
  \hat{a}^\dagger\vert n\rangle =\sqrt{n+1}\vert n+1\rangle , \label{adaggeraction}\\
  \vert n\rangle &=& \frac{1}{\sqrt{n!}} \left(\hat{a}^\dagger\right)^n \vert 0\rangle,\; \; n=0,1,2,.... \label{nfromzero} 
\end{eqnarray}
The ground state (vacuum) is given by the solution of
\begin{equation}\label{vacuuumeqn}
 \hat{a}\vert 0\rangle =0.
\end{equation}
In position representation ($\{\vert x\rangle \}$), the explicit solution of ~\eqref{vacuuumeqn} reads
\begin{eqnarray}\label{phizero}
 \langle x\vert 0\rangle = \phi_0(x)=  \sqrt[4]{\kappa/(\pi\hbar \sigma)} \exp \left(-K(t)x^2\right) ,\\
 \mbox{with}\; K(t)= \frac{1}{2\hbar\sigma}(\kappa - im\dot{\sigma}).
\end{eqnarray}
 If the variance $\Delta\hat{\mathcal{O}}\vert_{\phi}$ of an observable $\hat{\mathcal{O}}$ on the normalized state $\phi$ is given by 
 \begin{equation}
  \Delta\hat{\mathcal{O}}\vert_{\phi}=\sqrt{\langle \hat{\mathcal{O}}^2\rangle_{\phi} - \langle \hat{\mathcal{O}}\rangle_{\phi}^2},
 \end{equation}
through the expectation value $\langle \hat{\mathcal{O}}\rangle_{\phi}= \langle \phi\vert \hat{\mathcal{O}}\vert \phi\rangle $, then we have the uncertainty relation
\begin{eqnarray}
 \Delta \hat{x}\Delta \hat{p}\vert_{\phi_0} = \frac{\hbar}{2}\left(1+ \frac{m^2\dot{\sigma}^2}{\kappa^2}\right)^{\frac{1}{2}}.
\end{eqnarray}
Moreover,  $\Delta\hat{x}$ and $\Delta\hat{p}$ satisfy the equation of an ellipse
\begin{equation}
 \frac{(\Delta\hat{x})^2}{s_1^2} + \frac{(\Delta\hat{p})^2}{s_2^2} =1,
\end{equation}
where 
\begin{equation}
 s_1^2=\frac{\hbar\sigma}{\kappa}=\frac{\hbar}{\kappa_0}\alpha(t),\;\; s_2^2= \frac{\hbar\beta}{\sigma\kappa}=\frac{\hbar}{\kappa_0}\beta.
\end{equation}
The axis of $\Delta \hat{x}$ evolves with time, whereas the axis of $\Delta\hat{p}$ is time-independent. The semi-major axis and the semi-minor axis of the ellipse are interchanged with the time evolution.  The ellipse becomes a circle for $\sigma^2=\beta$  (the trivial case of a time-independent system), for which the ground state is indeed a coherent state (CS). However, in general, the state is a squeezed CS. 
\section{Solution of the time-dependent Schr\"{o}dinger equation}
If $\phi_n(x,t)$ is an eigenstate of $\hat{\mathcal{I}}$, then 
$\psi_n(x,t)=\phi_n(x,t)\exp(i\theta_n(t))$ will satisfy the TDSE ~\eqref{schrodinger} for some $\theta_n(t)$, which is given by ~\eqref{thetandot}. 
Let us express the phase factor ($\theta_n$) as a sum of the geometric ($\theta_n^g$) and dynamical ($\theta_n^d$) phases, i.e.,  $\theta_n=\theta^d_n + \theta_n^g$. Using ~\eqref{thetandot}, we get
\begin{eqnarray}
 \hbar\dot{\theta}_n^d &=& -\langle n \vert \hat{H}\vert n\rangle, \label{thetad}\\
 \hbar\dot{\theta}_n^g &=& i\hbar \langle n\vert \frac{\partial}{\partial t}\vert n\rangle. \label{geometricphase} 
\end{eqnarray}
To obtain the matrix element 
$
 h_{nn}=\langle n \vert \hat{H}\vert n\rangle
$,
it is convenient to rewrite ~\eqref{hamiltonian} in terms of $\hat{a}$ and $\hat{a}^\dagger$. In particular,
\begin{equation}\label{hina}
 \hat{H}= \frac{\hbar}{2m\omega}\left[ k_\alpha^2 \hat{a}^2+ (k^*_\alpha)^2 (\hat{a}^\dagger)^2 + \vert k_\alpha\vert^2(\hat{a}^\dagger\hat{a}+\hat{a}\hat{a}^\dagger)\right],\;\;
 \mbox{with} \; k_\alpha=\kappa + im\dot{\sigma}.
\end{equation}
Using ~\eqref{hina} in ~\eqref{thetad}, and utilizing  ~\eqref{adaggeraction}, we get
\begin{equation}\label{hnnform}
 h_{nn}= \frac{\hbar \beta}{2\kappa_0}(n+1/2)\dot{\mu}.
\end{equation}
Using ~\eqref{hnnform} in ~\eqref{thetad} we can determine the dynamical phases $\theta_n^d$, which reads
\begin{equation}\label{thetazerod}
 \theta_n^d =\int_0^t \dot{\theta}_n^d(\tau)d\tau= - \frac{\beta}{2\kappa_0}(n+1/2)\mu(t). 
\end{equation}
To calculate $\theta_n^g$ from the equation ~\eqref{geometricphase}, we first note that
\begin{equation}\label{adaggerderivative} 
 \frac{\partial \hat{a}^\dagger}{\partial t} = \Lambda_1 \hat{a} + \Lambda_2 \hat{a}^\dagger ,
\end{equation}
where
\begin{eqnarray}
 \Lambda_1 = -\frac{1}{2\kappa^2}\frac{d}{dt}(\kappa\kappa_\alpha),\;\;
 \Lambda_2 = \frac{\dot{\kappa}}{\kappa}- \frac{1}{2\kappa^2}\frac{d}{dt}(\kappa\kappa_\alpha^*).
\end{eqnarray}
On the other hand, using ~\eqref{Iinaadagger} and ~\eqref{vacuuumeqn}, we can write
\begin{equation}
 \frac{\partial \hat{\mathcal{I}}}{\partial t} \vert 0\rangle = \left(\frac{1}{2}\hbar\omega -\hat{\mathcal{I}}\right) \frac{\partial \vert 0\rangle}{\partial t}.
\end{equation}
Moreover, using the explicit form ~\eqref{phizero}, we get
\begin{equation}\label{thetazerog}
 \langle 0\vert \frac{\partial}{\partial t}\vert 0\rangle = \frac{\dot{\zeta}}{\zeta} -\frac{\dot{\Lambda}}{4\Lambda_r},
 \end{equation}
 with
 \begin{equation}
 \zeta = \sqrt[4]{\kappa/(\pi\hbar\sigma)},\;\; \Lambda= \frac{1}{2\hbar\sigma}(\kappa -im\dot{\sigma}),\;\; \Lambda_r=\operatorname{Re}(\Lambda).
\end{equation}
Integrating ~\eqref{thetazerog}, we can easily obtain $\theta_0^g$. For the general $\theta_n^g$, one can use ~\eqref{adaggerderivative}, along with ~\eqref{adaggeraction} and ~\eqref{nfromzero}  in ~\eqref{geometricphase} to get
\begin{eqnarray}\label{ndeltn}
 \langle n\vert \frac{\partial}{\partial t}\vert n\rangle = \frac{\dot{\zeta}}{\zeta} - \frac{\dot{\Lambda_r}}{4\Lambda_r} +   \frac{n\dot{\sigma}}{2\sigma} -\frac{i}{2\kappa_0}(n+1)m\dot{\sigma}^2 + \frac{i}{4\kappa_0}\frac{d}{dt}(m\sigma\dot{\sigma}).
\end{eqnarray}
Using ~\eqref{ndeltn} in ~\eqref{geometricphase} and integrating, we get the explicit form of the geometric phase, which reads
\begin{equation}\label{geometricphasefinal}
 \theta_n^g=-\frac{1}{2\kappa_0}(n+1/2)\gamma + \frac{i}{4}\ln(2\sigma^{2n}/\pi) +\frac{1}{2}(n+1)\tan^{-1}(\gamma /\kappa_0).
\end{equation}
Using ~\eqref{thetazerod} and ~\eqref{geometricphasefinal}, we can write the total phase as
\begin{equation}
 \theta_n (t) = -\frac{\gamma_0}{2\kappa_0}(n+1/2) + \frac{i}{4} \ln (2\sigma^{2n}/\pi) + \frac{1}{2}(n+1) \tan^{-1} (\gamma/\kappa_0).
\end{equation}
Thus we can write the general time-dependent states of the system by $\psi_n(x,t)=\phi_n(x,t)e^{i\theta_n(t)}$. For example, the ground state reads
\begin{equation}\label{tdemgroundstate}
 \psi_0(x,t)= \sqrt[4]{\kappa_0/(2\hbar\sigma^2)} \exp\left(  -Kx^2 -\frac{i\gamma_0}{4\kappa_0}+\frac{i}{2}\tan^{-1}(\gamma/\kappa_0)\right).
\end{equation}

\section{Wigner distribution}
The characteristic function $\hat{M}(\tilde{\tau},\tilde{\theta})$ of a random variable  $X=(\hat{x},\hat{p})$ is defined by $\hat{M}(\tilde{\tau},\tilde{\theta})= e^{i\mbox{Trace}(\tilde{t}^T X)},$ with the parameter $\tilde{t}=(\tilde{\tau},\tilde{\theta})$.
The Fourier transformation of the expectation value of the characteristic function ($\hat{M}$) is known as the Wigner quasiprobability distribution (WQD) \cite{wqd1,wqd2,wqd3}. 
In particular, WQD is related to the quantum wave function, which is obtained from the Schr\"{o}dinger equation, through the following integral transform \cite{wigner5,wigner6,wigner7,wqd1,wqd2,wqd3}.
\begin{equation}\label{wignerdefn}
 W(x,p,t)= \int_{-\infty}^{\infty} \psi^*(x+\frac{x'}{2})\psi(x-\frac{x'}{2})e^{\frac{ipx'}{\hbar}}dx'.
\end{equation}
We would now like to turn to a wave function which is composed of two  wave packets of the form ~\eqref{tdemgroundstate}, which we write as
\begin{equation}\label{psicomposite}
 \psi_T(x,t)= \frac{1}{\sqrt{2}}[\psi_0(x-x_0,t)+\psi_0(x+x_0,t)].
\end{equation}
This is thus one portion of the wave function centered at $x_0$, and a second portion centered at $-x_0$.
The probability density ($\rho_T=\psi_T^*\psi_T$) for the composite system ~\eqref{psicomposite} reads
\begin{equation}\label{rhocomposite}
 \rho_T(x,t)=\frac{1}{2}\sqrt{\kappa_0/(2\hbar\sigma^2)}[e^{-2k_r(x-x_0)^2} + e^{-2k_r(x+x_0)^2} + 2e^{-2k_r(x^2+x_0^2)}\cos(4k_ix_0x)],
\end{equation}
where $k_r(t) = \operatorname{Re}(K)$ and  $k_i(t)=\operatorname{Im}(K)$.\\
WQD ~\eqref{wignerdefn} for the composite system ~\eqref{psicomposite} reads
\begin{eqnarray}\label{wqdforcomposite}
 W(x,p,t)=\sqrt{\frac{\pi}{2}}[ e^{-\frac{2\vert K\vert^2 }{k_r}(x-x_0)^2 -\frac{2k_i}{\hbar k_r}p(x-x_0)}+ 
 e^{-\frac{2\vert K\vert^2 }{k_r}(x+x_0)^2 -\frac{2k_i}{\hbar k_r}p(x+x_0)}\nonumber \\
 +
 2 e^{-\frac{2\vert K\vert^2 }{k_r}x^2 -\frac{2k_i}{\hbar k_r}px} \cos(2x_0 p/\hbar)]e^{-\frac{p^2}{2\hbar^2 k_r}}.
\end{eqnarray}
It is worth noting that $k_i=0$ implies $\beta=\gamma=0$ and $\alpha=\alpha_0$ constant. In other words, $K\in \mathbb{R}$ corresponds to the constant mass system ($\hat{\mathcal{I}}=\alpha_0\hat{p}^2$). Using $k_i=0$ in ~\eqref{wqdforcomposite}, one can verify that it is consistent with the well known WQD corresponding to the constant mass cat-state \cite{wigner5}.\\
A toy model is illustrated graphically in the next section.

\section{Toy Model}
\begin{center}
    \begin{figure}[!h]
    \centering\includegraphics[totalheight=4cm]{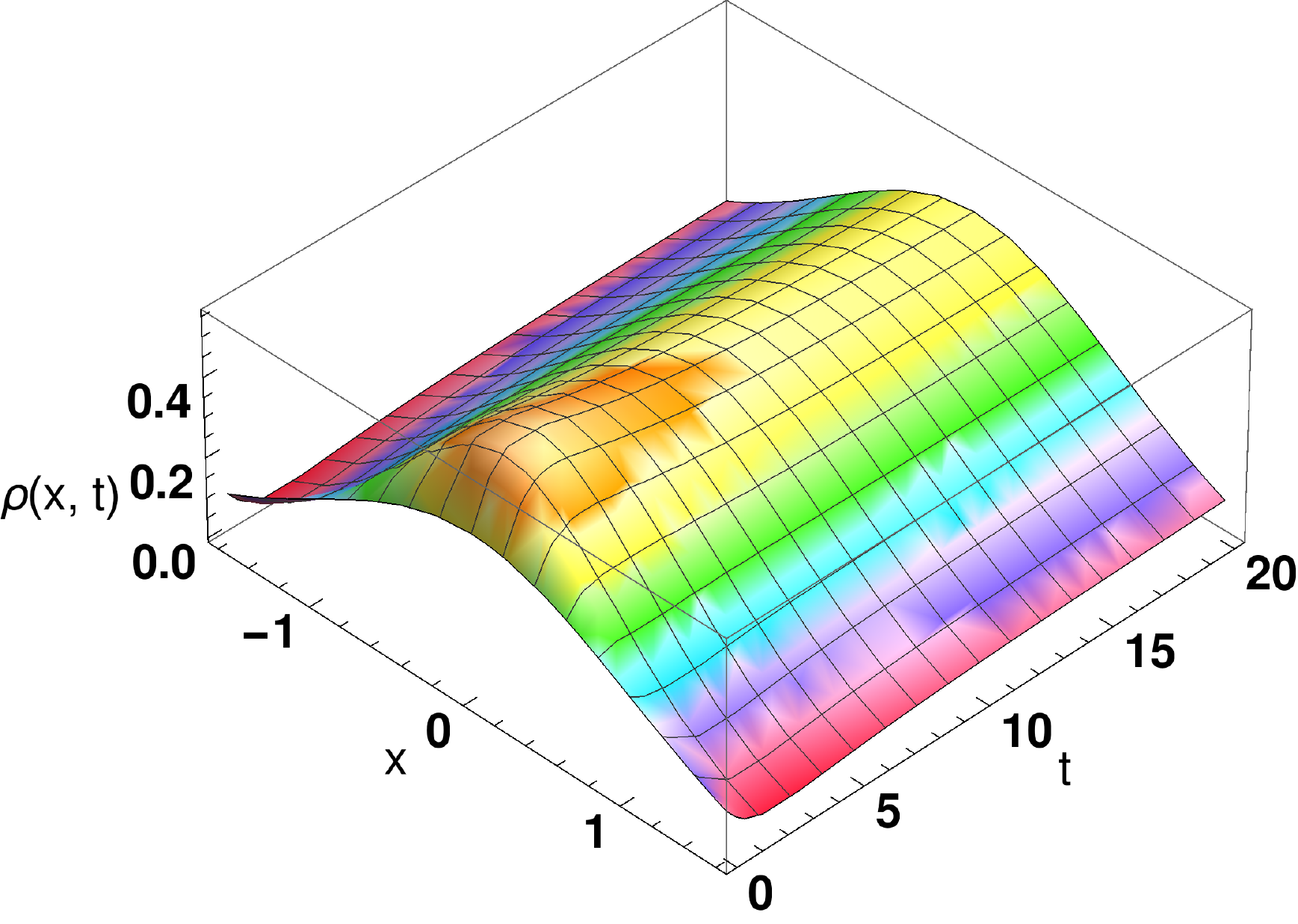}\\
\caption{\textbf{probability density $\rho(x,t)=\psi_0^*\psi_0$ .}}\label{sampleFig1}
    \end{figure}
    \end{center}

\begin{center}
    \begin{figure}[!h]
    \centering\includegraphics[totalheight=4cm]{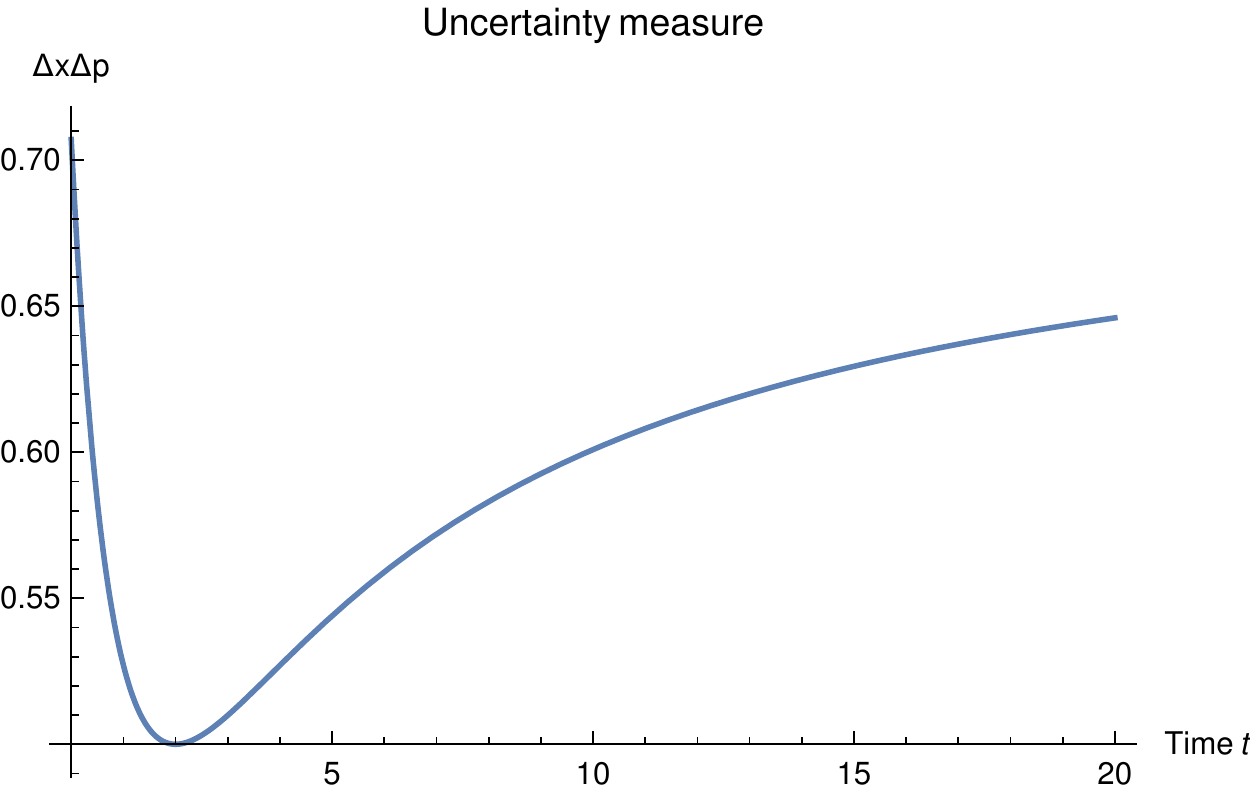}\\
\caption{\textbf{Variance with respect to time. Clearly $\Delta\hat{x}\Delta\hat{p}\ge\frac{1}{2}$.}}\label{sampleFig2}
    \end{figure}
\end{center}

\begin{center}
\begin{figure}[!h]
\centering\includegraphics[totalheight=4cm]{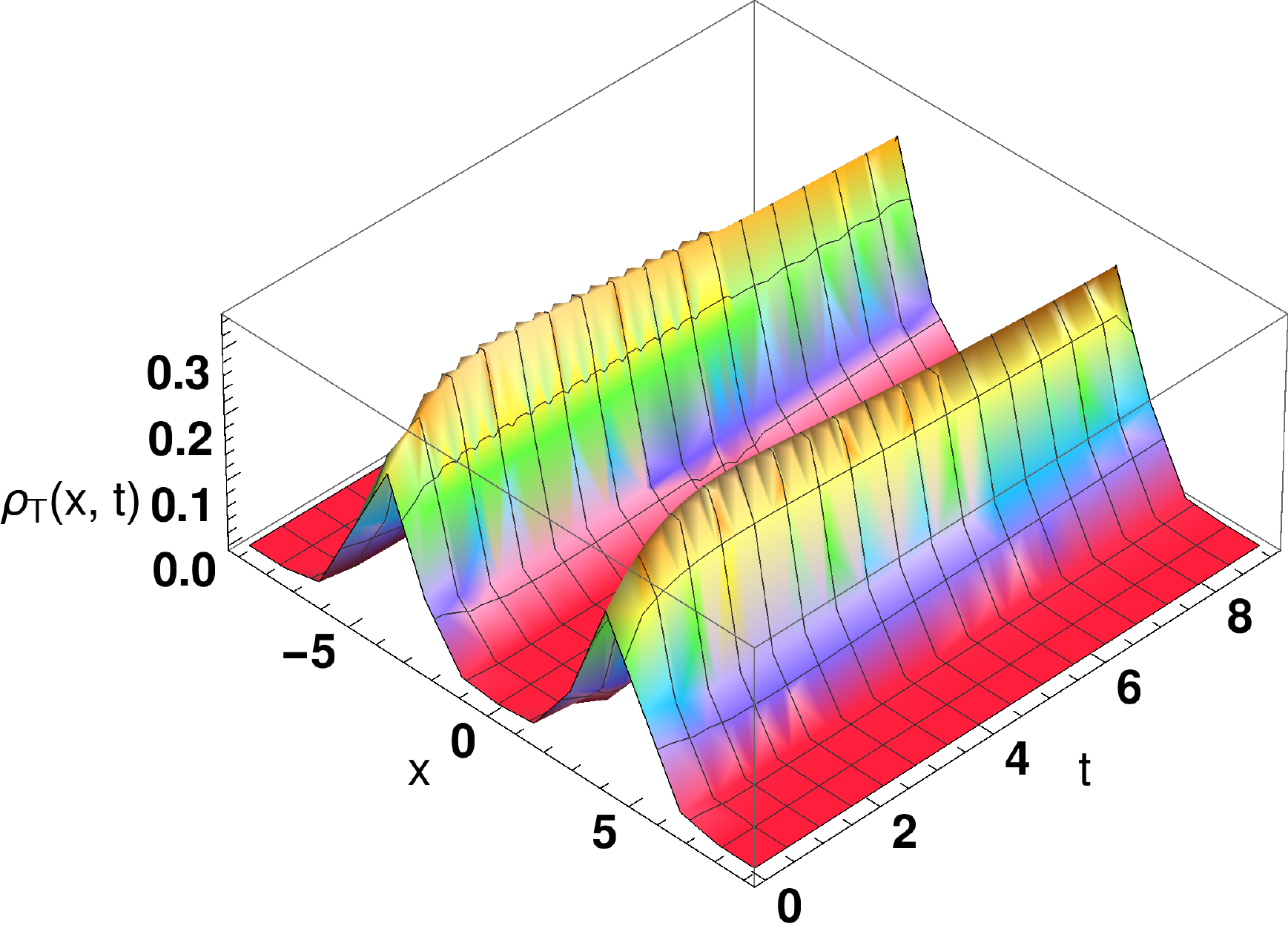}\\
\caption{\textbf{Probability density ($\rho_T=\vert \psi_T\vert^2$)for the composite system (Considering $x_0=4$) }}\label{sampleFig3}
    \end{figure}
\end{center}

\begin{figure}[!h]
\begin{minipage}[t]{0.5\textwidth}
\centering\includegraphics[width=\textwidth]{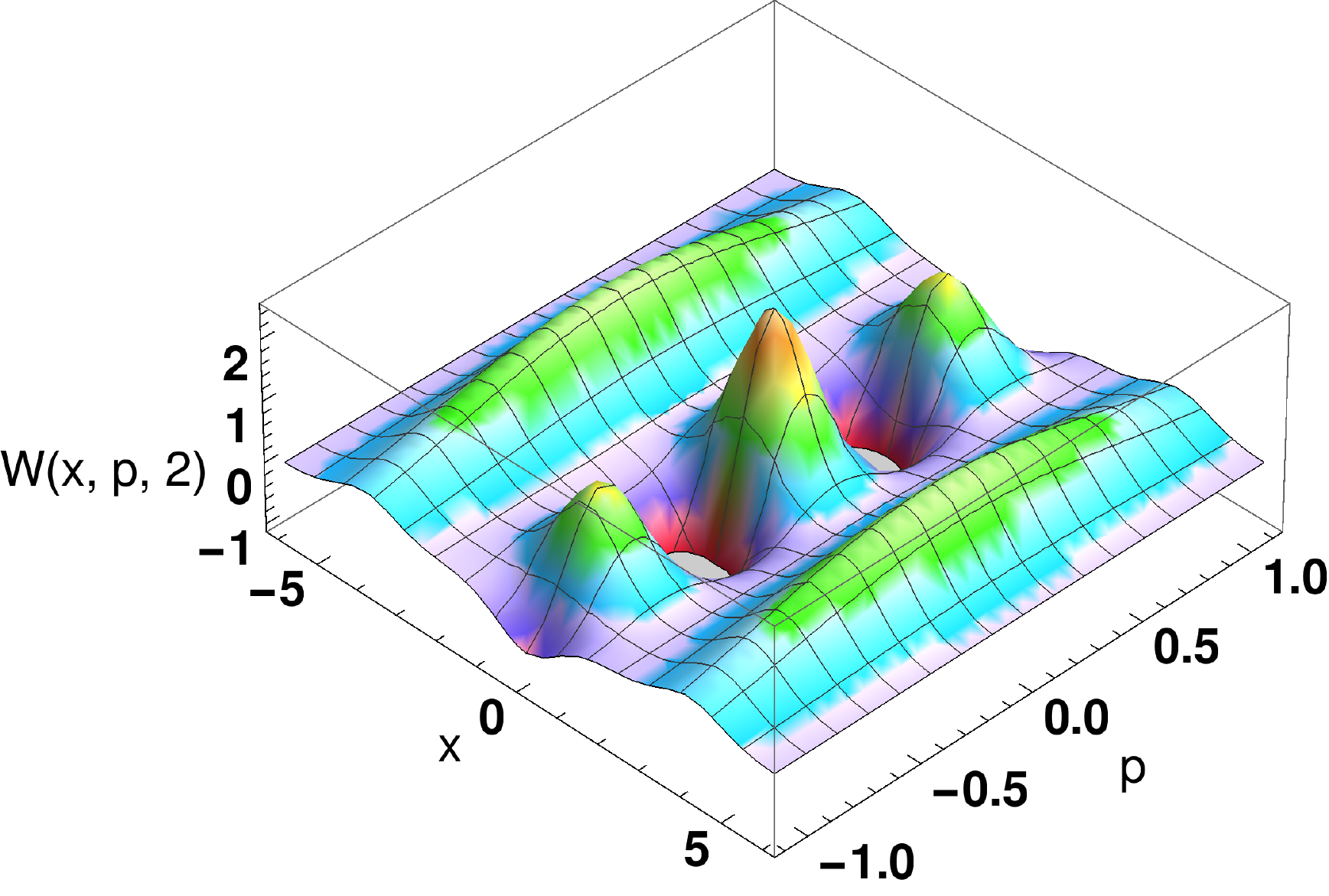}\\
{\footnotesize $t=2$}
\end{minipage}\hfill
\begin{minipage}[t]{0.5\textwidth}
\centering\includegraphics[width=\textwidth]{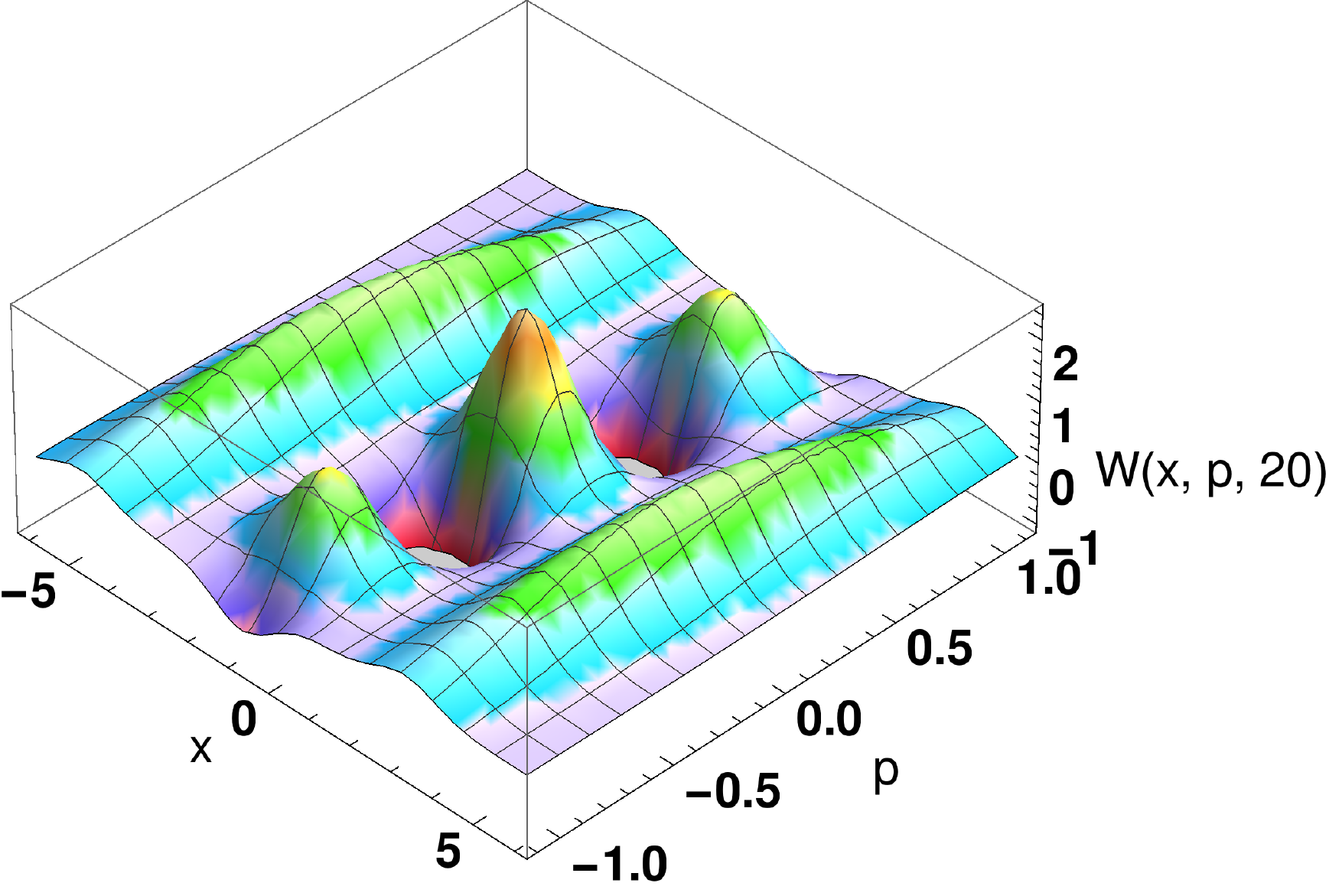}\\
{\footnotesize $t=20$}
\end{minipage}
\caption{\textcolor{blue}{\textbf{ Wigner distributions show that the correlation at the origin remains forever. }}}\label{sampleFig4}
\end{figure}

As a toy model, we shall consider the TDEM 
\begin{equation}\label{mt}
 m(t)=m_0(1+bt)^2,
\end{equation}
which appears to be effective in modeling the transport phenomena in the electronic band structure \cite{toy1}. In particular, ~\eqref{mt} is incorporated to account for phenomena such as electron-phonon scattering that orchestrate relaxation in charge carrier energy observed in nanostructures \cite{toy1,toy2}. 
~\eqref{mt} implies
\begin{equation}\label{mut}
 \mu(t)= \frac{t}{m_0(1+bt)}.
\end{equation}
One can retrieve the constant mass ($m_0$) case by setting the parameter $b\to 0$.
For the visual illustration, let us choose the following convenient values of the parameters.
\begin{equation}\label{choiceparameter}
 m_0=1, \;  \beta=1,\; \alpha_0=2,\; \gamma_0=1.
\end{equation}
Using the value ~\eqref{choiceparameter} of the parameters, we get
\begin{eqnarray}
\gamma=1-\mu,\; \alpha= 1+\gamma^2,\\
 k_i= \gamma k_r,\;k_r = \frac{1}{2\hbar(1+\gamma^2)} .
\end{eqnarray}
The uncertainty measure (through the variance of the observables) reads as
\begin{equation}
 \Delta\hat{x}\Delta\hat{p}=\frac{\hbar}{2}\sqrt{1+ \gamma^2}.
\end{equation}
From nowon we shall consider $b=0.5$ and $\hbar=1$ throughout our discussion.\\
For the choice~\eqref{choiceparameter}, the FIG.~\ref{sampleFig1} represents the probability density corresponding to $\psi_0(x,t)$.
  The uncertainty measure ($\Delta\hat{x}\Delta\hat{p}$)  is on the FIG.~\ref{sampleFig2}, which indicates that the state is a squeezed CS.  
The composite system ~\eqref{psicomposite} is considered with $x_0=4$.
The probability density ($\rho_T= \psi_T^*\psi_T$) corresponding to ~\eqref{psicomposite} is shown in FIG.~\ref{sampleFig3}.
 From FIG.~\ref{sampleFig4}, we can envisage the time variation of WQD corresponding to the composite system $\psi_T$. 
\begin{figure}[!h]
\begin{minipage}[t]{0.5\textwidth}
\centering\includegraphics[width=\textwidth]{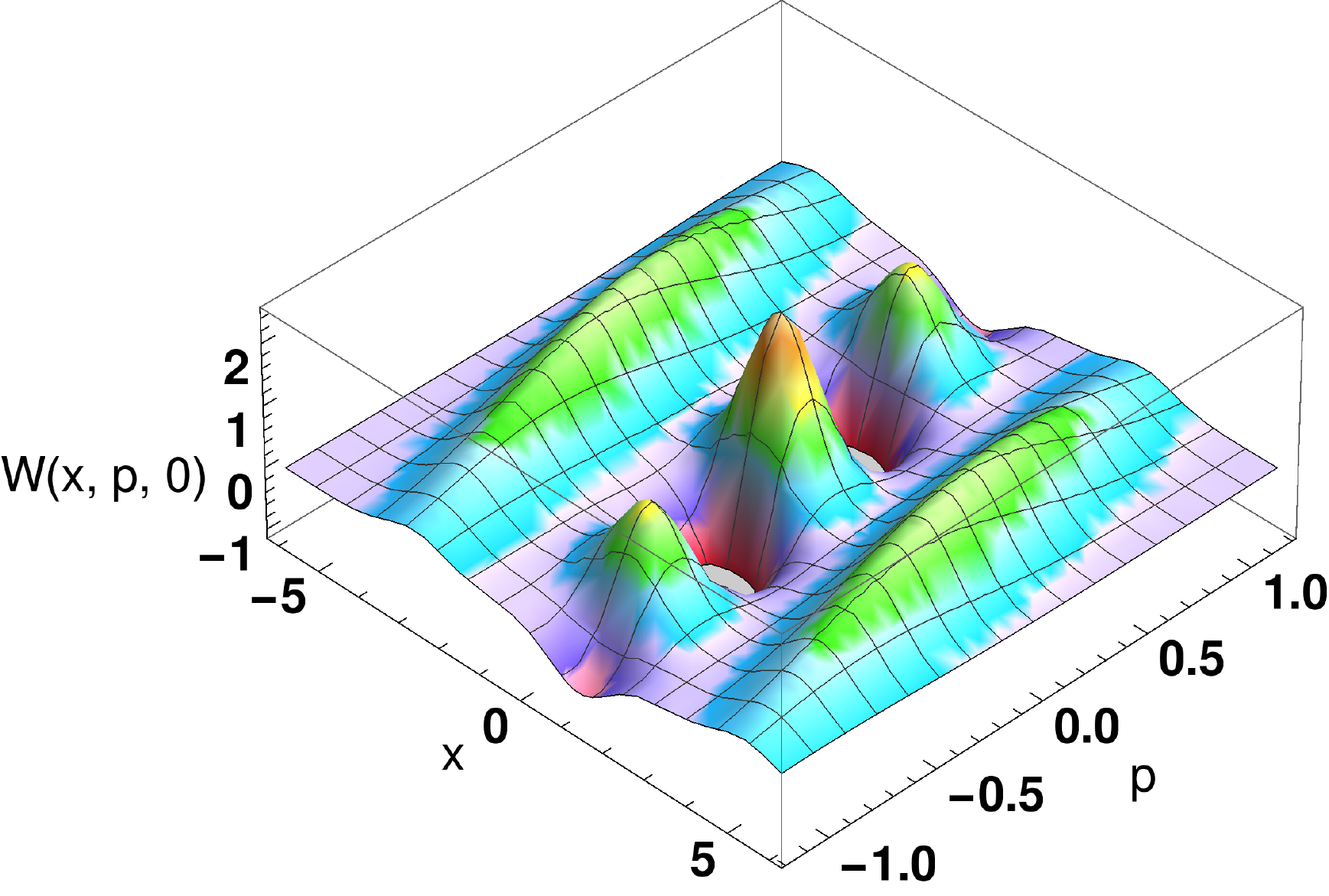}\\
{\footnotesize $t=0$}
\end{minipage}\hfill
\begin{minipage}[t]{0.5\textwidth}
\centering\includegraphics[width=\textwidth]{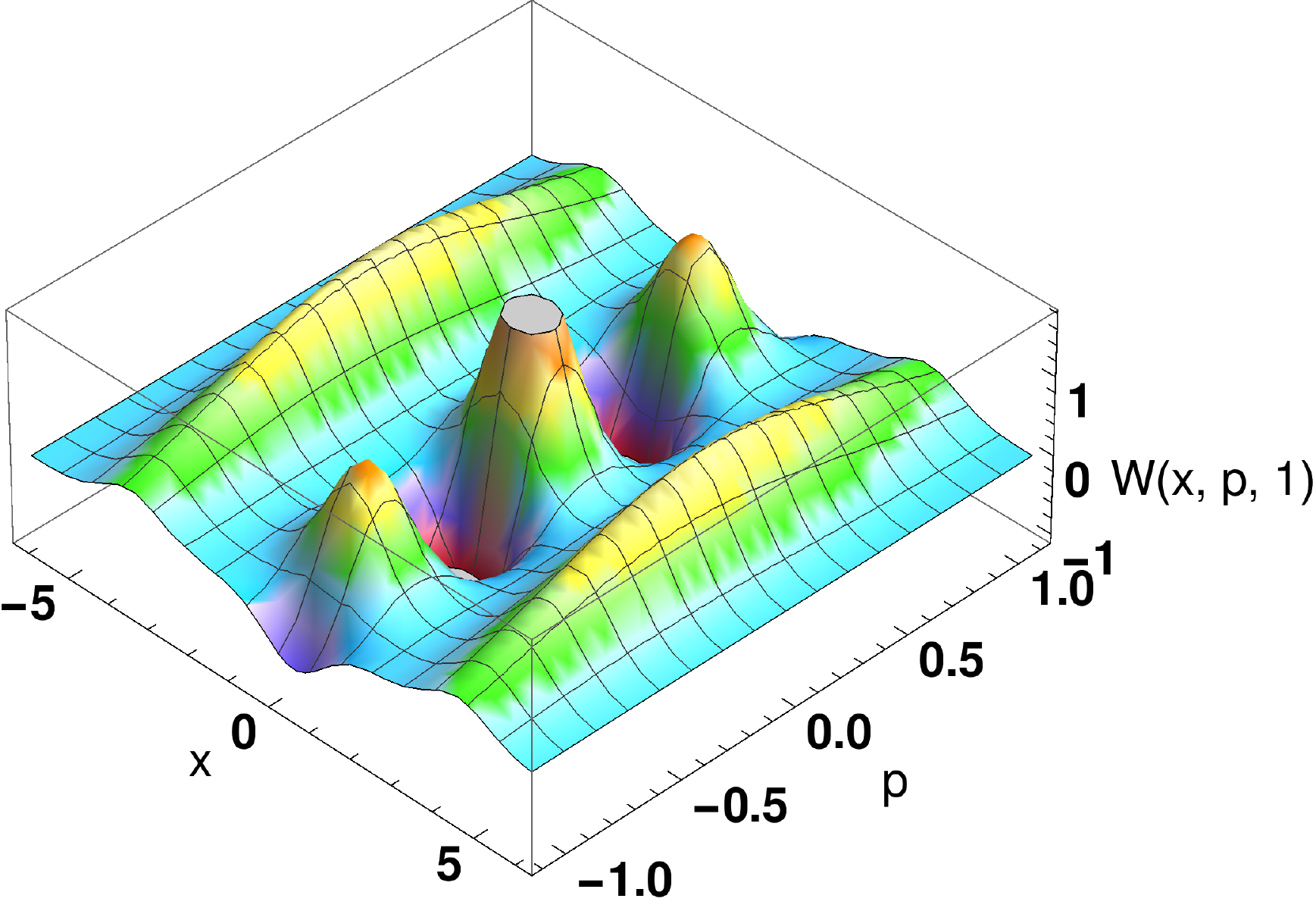}\\
{\footnotesize $t=1$}
\end{minipage}
\end{figure}
Since $\mu(t) \to 1/(m_0b)$, the amplitude of WQD remains almost unaltered for large $t$. $W(x,p,2)$ and $W(x,p,20)$ on FIG.~\ref{sampleFig4} are showing this feature. For small $t$, the amplitude changes notably, which can be seen from the figure of $W(x,p,1)$.
FIG.~\ref{sampleFig4} depicts  that the entanglement at the origin remains forever. In other words, the particles are never separated. They leave their trace at the origin of the phase-space. Since our mass function is a monotone increasing function in time, we expect that the quantum effect will goes off after some time. However, contrary to this, we have seen that the quantumness of the system remains intact for all time.
\section{Conclusions}
Our study uncovers two aspects. Firstly, we have developed the scope of  Lewis-Riesenfeld (LR)- invariant operators for a quantum system of free-particle (FP) with time-dependent(TD) effective mass (EM). Secondly, we have studied the entanglement behavior of such a bipartite system. \\
On the way of diagonalizing the quadratic LR-invariant operator (LRIO), we have obtained the constraint on the free parameters so that LRIOs can be diagonalized with a unitary transformation (symplectic group Sp($4,\mathbb{R}$)). It turns out that the LRIOs can be factorized with the annihilation and creation operators, which enable us to obtain the eigenfunctions and eigenvalues of the LRIOs in closed form.  Moreover, we have shown that the ground state eigenfunctions of the LRIOs are indeed squeezed coherent states. Both the geometric and dynamical phase (time-dependent) corresponding to the general $n^{th}$ state have been determined.  Thus we have completely solved the TD-Schr\"{o}dinger equation for an FP with general TDEM. \\
In the next part, we have considered a bi-partite system, which is composed of two wave packets, one portion of which is centered at $x_0$, and a second portion is centered at $-x_0$. 
The Wigner quasiprobability distribution for such a spatially separated state is constructed. It turns out that,  the particles are never separated, even for a TDEM.  It is often argued that the quantumness of the system will be diminished for an object with a large mass. Since our toy model TDEM is a monotone increasing function of time, one can expect that the quantumness of the system will go off after a long time. However, this is not observed in our phase-space distribution. Therefore, the long-standing expectation of achieving the classical theory as a limiting case of quantum theory might not be materialized for such a straightforward phase-space distribution approach. Rather, it demands a fresh study on the interpretation of the quantum world and classical world.
\section{Acknowledgement}
The authors AC and MJ are grateful to Brahmananda Keshab Chandra College for the hospitality. We are grateful to the anonymous referee for the fruitful suggestions.
\section{Data Availability Statement}
The present manuscript has no associated data.

\end{document}